\documentstyle[preprint,eqsecnum,aps]{revtex}

\begin{document}

\hoffset-1cm

\draft
\preprint{TRP-95-12}

\title{
   Transverse momentum dependence of\\
   Hanbury-Brown/Twiss correlation radii
}

\author{Urs Achim Wiedemann, Pierre Scotto and Ulrich Heinz}

\address{
   Institut f\"ur Theoretische Physik, Universit\"at Regensburg,\\
   D-93040 Regensburg, Germany
}
\date{\today}

\maketitle

\begin{abstract}
\baselineskip18pt
  The transverse momentum dependence of Hanbury-Brown/Twiss (HBT)
  interferometry radii for 2-body correlation functions provides
  experimental access to the collective dynamics in heavy-ion
  collisions. We present an analyti\-cal approximation scheme for
  these HBT radii which combines the recently derived
  model-independent expressions with an approximate determination of
  the saddle point of the emission function. The method is illustrated
  for a longitudinally boost-invariant hydrodynamical model of a heavy
  ion collision with freeze-out on a sharp hypersurface. The
  analytical approximation converges rapidly to the width of the
  numerically computed correlation function and reproduces correctly
  its exact transverse momentum dependence. However, higher order
  corrections within our approximation scheme are essential, and the
  previously published lowest order results with simple $m_\perp$
  scaling behaviour are quantitatively and qualitatively unreliable.
\end{abstract}


\section{Introduction}\label{sec1}

While the total energy involved in a heavy ion collision can be
measured directly by particle calorimeters, an analogous direct
measurement of the locally reached energy density does not exist. An
indirect determination of the size of the interaction region is
possible through Hanbury-Brown/Twiss (HBT) intensity interferometry
\cite{BGJ90}. However, the interpretation of the measured correlations
between the produced particles is in general model-dependent, and a
considerable amount of theoretical effort has recently been spent
on the question to what extent this intrinsic model dependence can be
reduced by a refined analysis \cite{MSH92,CL94,CSH95,CNH95,S95,AS95}.

Compared to HBT interferometry on stars, the situation in heavy-ion
collisions is complicated by the finite lifetime and the strong
dynamical evolution of the particle emitting source
\cite{KP72,P86,MS88,PCZ90}. As a result the 2-body correlation function is
in general characterized by different width parameters (``HBT radii")
if the relative momentum between the two identical particles points in
different directions \cite{P86}. Furthermore, the dynamical expansion
of the source leads to correlations between the momenta of the emitted
particles and their emission point (so-called $x-K$ correlations)
which in turn generate a characteristic dependence of the HBT radii on
the total pair momentum \cite{P86}.

In this paper we discuss this last issue in some detail. Our work was
motivated by the simple scaling laws for the HBT radii as a function
of the transverse mass $m_\perp$ of the particle pair which were
proposed in Refs.~\cite{MS88,CL94,S95,AS95,HB95}. These predictions are
based on simple models for the emission function, and it is natural to
ask which of the predicted features are independent of the model and
which are not. For example, the model of Ref.~\cite{AS95} makes
definite assumptions about the shape of the longitudinal and
transverse expansion flow profiles whose influence on the result for
the correlation function, in particular on the functional form of the
$m_\perp$-dependence of the correlation radii, is not known. The work
reported here provides an answer to this question within a certain
class of models. More importantly, however, we discovered on our way
that the simple scaling laws of Refs.~\cite{MS88,CL94,S95,AS95} are
based on approximations which are quantitatively unreliable and in
some typical cases give even qualitatively misleading results. We will
show, for example, that a fit of the exact correlation function with
the simple analytical expression from Ref.~\cite{MS88} for the
longitudinal HBT radius $R_l$ gives an estimate for the decoupling
time which is too large by a factor $\simeq 2$. (This was recently
also pointed out in \cite{HB95}.) The effect of the finite duration of
particle emission on the difference between the HBT radii in ``out"
and ``side" directions, $R_o^2-R_s^2$ which, as we will show, is also
present in models with sharp freeze-out along a proper-time
hypersurface, is completely missed by the approximations used in
Ref.~\cite{AS95}, giving rise to a qualitatively wrong $m_\perp$
dependence of the ``out"-correlator. We conclude from our findings
that a quantitative analysis of the $m_\perp$-dependence of HBT radii
necessitates a numerical evaluation of the theoretical expressions for
the correlation radii, or requires at least a much more sophisticated
analytical approximation scheme than so far employed.

We begin by shortly reviewing the general approach and defining our
notation. We restrict our discussion to two-particle correlations. We
start from the usual abstraction of the collision region as an
assembly of classical boson emitting sources in a certain space-time
region \cite{BGJ90,GKW79}. Their Wigner transform \cite{S73,P84,CH94}
defines the emission function $S(x,p)$ which describes the probability
that a particle with on-shell momentum $p$ ($p^2=m^2$) is emitted from
the space-time point $x$. The emission function determines the single
particle momentum spectrum $P_1({\bf p}) = E_p\, dN/d^3p = \int d^4x\,
S(x,p)$ as well as the HBT two-particle correlation function $C({\bf
K},{\bf q})$ where ${\bf K} = {1\over 2}\left({\bf p}_1 + {\bf p}_2
\right) = ({\bf K}_\perp, K_L)$ is the average momentum of the two
particles and ${\bf q} = {\bf p}_1 -{\bf p}_2$ their relative
momentum. In the plane wave approximation the latter is given in terms
of the 1- and 2-particle distributions $P_1({\bf p})$, $P_2({\bf
p}_1,{\bf p}_2)$ and the average number $\langle N \rangle$ ($\langle
N(N-1) \rangle$) of particles (particle pairs) produced in the
reaction as \cite{GKW79,S73,CH94}
 \begin{equation}
   C({\bf p}_1,{\bf p}_2) =
   {\langle N \rangle^2 \over \langle N(N-1) \rangle}
   {P_2({\bf p}_1,{\bf p}_2)\over P_1({\bf p}_1) P_1({\bf p}_2)}
   = 1 + {\left\vert \int d^4x\, S(x,K)\,
   e^{iq{\cdot}x}\right\vert^2 \over
   \int d^4x\, S(x,p_1)\ \int d^4x\, S(x,p_2) } \, .
 \label{1}
 \end{equation}
In the numerator of the last expression we have introduced off-shell
momentum 4-vectors $K$ and $q$ for the total and relative momentum of
the particles in the pair by defining $K^0 = {1\over 2} (E_1 + E_2)$
and $q^0 = E_1 -E_2$ where $E_i = \sqrt{m^2 + {\bf p}_i^2}$. A popular
approximation whose accuracy was studied quantitatively in \cite{CSH95}
consists of putting $K$ on-shell, $K^0 \simeq E_K = \sqrt{m^2+{\bf
K}^2}$, and setting $p_1{=}p_2{=}K$ in the denominator:
 \begin{equation}
   C({\bf K},{\bf q}) \simeq 1 +  {\left\vert \int d^4x\, S(x,K)\,
   e^{iq{\cdot}x}\right\vert^2 \over
   \left\vert \int d^4x\, S(x,K) \right\vert^2 } \, .
 \label{2}
 \end{equation}
The aim of HBT interferometry is to obtain from the measured
correlation function $C({\bf K},{\bf q})$ information about $S(x,K)$
which in turn should characterize the size of the interaction region.
One usually compares $C({\bf K},{\bf q})$ to experimental data by
taking recourse to the Gaussian approximation \cite{CSH95}
 \begin{equation}
   C({\bf K},{\bf q}) \simeq 1 + \exp\left[ - q_s^2 R_s^2({\bf K})
   - q_o^2 R_o^2({\bf K}) - q_l^2 R_l^2({\bf K})
   - 2 q_l q_o R_{lo}^2({\bf K})\right] \, ,
 \label{3}
 \end{equation}
which is valid for azimuthally symmetric sources. Here $q_l$, $q_o$,
$q_s$ denote the spatial components of ${\bf q}$ in the beam direction
(``longitudinal" or $z$-direction), parallel to the transverse
components ${\bf K}_\perp$ of ${\bf K}$ (``out" or $x$-direction), and
in the remaining third cartesian direction (``side" or $y$-direction),
respectively. In practice the last term in (\ref{3}) which mixes the
components $q_o$ and $q_l$ has usually been neglected because its
existence was only recently pointed out \cite{CSH95} and confirmed
experimentally \cite{A95}.

The main purpose of the present work is to investigate the transverse
momentum dependence of the HBT correlation radii $R_i^2({\bf K})$ in
(\ref{3}). As already mentioned it reflects the $x-K$-correlations in
the emission function $S(x,K)$ generated by collective expansion of
the source, information that can not be obtained from
${\bf K}$-averaged correlation radii. Our analysis will be based on a
specific model for the emission function, but the method is general
and can later be combined with a comprehensive investigation of the
model dependence of 2-particle correlations. In the present paper we
will compare two different methods to calculate the $R_i^2({\bf K})$:
the first consists of fitting the numerically determined HBT
correlation function $C({\bf K},{\bf q})$ to the form (\ref{3}) in
such a way that its half width is correctly reproduced, while in the
second approach we will use the following model-independent
expressions \cite{HB95,CSH95,fn1}:
 \begin{eqnarray}   R_s^2 &=& \langle y^2\rangle,
 {\nonumber} \\
   R_o^2 &=&  \langle (x-{\beta}_{\perp}t)^2\rangle
                 - {\langle (x-{\beta}_{\perp}t)\rangle}^2,
 {\nonumber} \\
   R_l^2 &=& \langle (z-{\beta}_{L}t)^2\rangle
                 - {\langle (z-{\beta}_{L}t)\rangle}^2,
 {\nonumber} \\
   R_{lo}^2 &=& \langle (x-{\beta}_{\perp}t)(z-{\beta}_{L}t)\rangle
                 -\langle (x-{\beta}_{\perp}t)\rangle
                  \langle(z-{\beta}_{L}t)\rangle \, ,
 \label{4}
 \end{eqnarray}
which are guaranteed to yield the correct Gaussian curvature of the
correlation function $C({\bf K},{\bf q})$ near ${\bf q}=0$
\cite{CSH95,CNH95,HB95}. In Eqs.~(\ref{4}) we defined ${\beta}_i =
2K_i/(E_1+E_2)\approx K_i/E_K$ and introduced the notation
 \begin{equation}
   \langle \xi\rangle = \langle \xi\rangle (K) =
   {\int d^4x\, \xi\, S(x,K)\over\int d^4x \, S(x,K)} \, .
 \label{5}
 \end{equation}
For Gaussian sources, the set of equations (\ref{4}) can be
alternatively derived by expanding (\ref{1}) for small relative
momenta ${\bf q}$ up to second order and re-exponentiating
\cite{CSH95,HB95}, or by making a Gaussian saddle point approximation
of the emission function around its maximum \cite{CNH95}. A more
general justification of Eqs.~(\ref{4}), which (as we will show)
remain rather accurate even for non-Gaussian sources, is based on
the following parametrization of the emission function:
 \begin{equation}
    S(x,K) = S(\bar{x},K)\,
    e^{-{1\over 2} (x-\bar{x})^{\mu} (x-\bar{x})^{\nu}
    B_{\mu\nu}({\bf K})} + \delta S(x,K)\, ,
 \label{5a}
 \end{equation}
where the saddle point $\bar x({\bf K})$ is defined as the point where
all first derivatives of $S(x,K)$ vanish. If one identifies
$B_{\mu\nu}$ with the tensor of second derivatives of $\ln
S(\bar{x},K)$, the first term on the r.h.s. of Eq.~(\ref{5a})
amounts to the Gaussian saddle point approximation around $\bar{x}$
which was used in \cite{CNH95,S95,AS95}. For non-Gaussian forms of
$S(x,K)$, the typical corrections to Eqs.~(\ref{4}), (\ref{5}) from
$\delta S(x,K)$ can be minimized by instead defining $B_{\mu\nu}({\bf
K})$ in terms of the variance of $S(x,K)$,
 \begin{equation}
 (B^{-1})_{\mu\nu} =  \langle x_{\mu}x_{\nu}\rangle
                 -\langle x_{\mu}\rangle \langle x_{\nu}\rangle \, .
 \label{5b}
 \end{equation}
which measures the width of $S(x,K)$. With this definition the Fourier
transform (\ref{2}) of the first term in (\ref{5a}) can be done
analytically and leads directly to the expressions (\ref{3}) and
(\ref{4}), with the expectation value (\ref{5}) defined in terms of
the {\em full} emission function $S(x,K)$. We will see that, in
contrast to other approximation schemes (see for example
\cite{S95,AS95}), the determination of the HBT radii via (\ref{4})
yields an accurate representation of the width of the exact
correlation function, independent of the validity of the Gaussian
saddle point approximation for $S(x,K)$.

Except for the few special cases for which the four-dimensional
Fourier transform of $S(x,K)$ is known exactly, an analytical
investigation of equation (\ref{1}) or (\ref{4}) must
involve a suitable approximation scheme. The existing analytical
calculations \cite{CL94,CSH95,S95,AS95} of (the ${\bf K}$-dependence
of) the HBT radii for particular emission functions $S(x,K)$ cover
certain limiting cases of parameter space only, and their validity has
not been checked numerically. In the present work, we investigate the
complete ${\bf K}$-dependence of the HBT radii by both analytical and
numerical methods without any restriction to limiting cases.

Our paper is organized as follows: In section \ref{sec2}, we review a
recently introduced simple model for the emission function
\cite{S95,AS95} which we will also study here. It may not be the most
realistic model for heavy-ion collisions, but it possesses a number of
essential physical features and allows for a controlled investigation
of our analytical approximation scheme and a direct comparison with
previously published approximations. In section \ref{sec3} we develop
a new analytical approximation scheme for the HBT radii (\ref{4}) of
this model. In Section \ref{sec4} we provide some intermediate results
for the numerical evaluation of the correlation function. In Section
\ref{sec5} we compare the analytical and numerical values for the HBT
radii. We find that the analytical approach works very well for a
linear transverse flow profile, if the approximation scheme (which
involves an expansion in powers of $\gamma_t m_\perp/T$,  $\gamma_t$
being the Lorentz factor associated with the transverse flow) is
carried to third non-leading order. For the ``out" and ``longitudinal"
radii the leading terms in this approximation scheme are found to be
insufficient. For parabolic transverse flow profiles our approximation
scheme and, {\em a fortiori}, all previously suggested simpler
approximations are found to fail completely. The physical consequences
of our findings are discussed in Section \ref{sec6}.

\section{A simple model for the emission function}\label{sec2}

For ease of comparison with published results in the literature, we
consider the emission function \cite{S95,AS95}
 \begin{equation}
   S(x,K) = {K{\cdot}n(x)\over ({2\pi})^3}\, e^{- K{\cdot}u/T}
   \, e^{-r^2/(2 R^2)} \, ,
 \label{6}
 \end{equation}
where the Boltzmann factor $e^{- K{\cdot}u/T}$ reflects the
assumed local thermal equilibration of a source with local
temperature $T(x)$ moving with hydrodynamic $4$-velocity $u_{\mu}(x)$.
We will take $T$ to be constant. We assume sharp freeze-out of the
particles from the thermalized fluid along a hypersurface $\Sigma(x)$,
and the 4-vector $n_{\mu}(x) = \int_{\Sigma} d^3\sigma_{\mu}(x')\,
{\delta}^{(4)}(x-x')$ denotes the normal-pointing freeze-out
hypersurface element. The factor $\exp\left[-r^2/(2 R^2) \right]$
imposes a Gaussian transverse density profile with geometric radius
$R$; the exponent can be interpreted in terms of an effective chemical
potential $\mu(r){=}- T\,r^2 /(2 R^2)$ which depends only on the
transverse coordinate $r{=}\sqrt{x^2+y^2}$.

For the flow velocity profile $u^\nu(x)$ we assume longitudinal boost
invariance by setting $v_L{=}z/t$, i.e. identifying the flow rapidity
$\eta_{\rm flow}{=} {1 \over 2} \ln[(1 + v_L)/(1-v_L)]$ with the
space-time rapidity $\eta{=} {1 \over 2} \ln[(t+z)/(t-z)]$. We can
thus parametrize $u^\nu(x)$ in the form
 \begin{equation}
 u_{\nu}(x) = ({\rm ch\,}\eta\, {\rm ch\,}\eta_t,\,
                 {\rm sh\,}\eta_t\, {x\over r},\,
                 {\rm sh\,}\eta_t\, {y\over r},\,
                 {\rm sh\,}\eta\, {\rm ch\,}\eta_t)\, ,
 \label{7}
 \end{equation}
where $\eta_t(r)$ is the transverse flow rapidity. The momentum $K$ is
parametrized in the usual way through the transverse mass
$m_\perp{=}\sqrt{m^2 + K_\perp^2}$ and the rapidity $Y$ as
$K_\nu{=}(m_\perp {\rm ch\,}Y, K_{\perp}, 0, m_\perp {\rm sh\,} Y)$.

Assuming sharp freeze-out at constant proper time ${\tau}_0$, the
freeze-out hypersurface is parametrized as $\Sigma(x){=}(\tau_0{\rm
ch\,} \eta, x, y, \tau_0{\rm sh\,}\eta)$, resulting in
$K{\cdot}n(x){=}m_\perp {\rm ch\,}(\eta -Y)\, \delta(\tau - \tau_0)$.
The emission function can thus be rewritten as
 \begin{equation}
   S(x,K) = {m_\perp {\rm ch\,}(\eta-Y) \over ({2\pi})^3}\,
            e^{- \left[ m_\perp {\rm ch\,}(\eta-Y)\, {\rm ch\,}\eta_t
               - K_\perp (x/r)\, {\rm sh\,}\eta_t \right]/T} \,
            e^{-r^2/(2 R^2)} \,
            \delta(\tau-\tau_0)\, .
 \label{8}
 \end{equation}
The $\delta$-function renders the $\tau$-integration in $d^4x{=}\tau\,
d\tau\, d\eta\,dx\,dy$ trivial, and only the integrations over
space-time rapidity $\eta$ and the transverse coordinates $x,y$ remain
to be done. Exploiting the boost invariance and infinite longitudinal
extension of our source we can simplify the structure of
Eqs.~(\ref{4}) by going to the longitudinally comoving system (LCMS).
In this frame $Y{=}0{=}\beta_L$.

\section{Analytical evaluation of the HBT radii}\label{sec3}

In this section we will discuss the model-independent expressions
(\ref{4}) for the HBT radii. Our analysis of these expressions
proceeds in two steps: First, the $\eta$-integration is done
analytically. We then find the approximate saddle point of the
resulting integrand in the $x$- and $y$-directions and carry out those
integrations by saddle point integration.

\subsection{Integration over $\eta$}\label{sec3A}

To begin with we note that by the symmetry of our infinite
boost-invariant emission function (\ref{8}) the expectation values of
all odd functions of $\eta$ vanish. Therefore, in this model the cross
term $R_{lo}$ is zero \cite{CSH95}.

Writing $t{=}\tau {\rm ch\,} \eta$, $z{=}\tau {\rm sh\,}\eta$ and
including the
${\rm ch\,}\eta$ prefactor in Eq.~(\ref{8}) (recall that we use the LCMS
frame where $\beta_L{=}Y{=}0$), the $\eta$-integrations in the
expectation values (\ref{4}) can be done analytically and expressed in
terms of modified Bessel functions $K_{\nu}(a) = \int_0^\infty d\eta\,
e^{-a\,{\rm ch\,}\eta}\, {\rm ch\,}(\nu\eta)$. Defining
 \begin{equation}
 G^{mn}(r) =
   \int_{-\infty}^\infty d\eta\, {\rm sh}^m\eta\, {\rm ch}^{n+1}\eta\,
   e^{-a(r)\,{\rm ch\,}\eta}
   \Bigg/
   \int_{-\infty}^\infty d\eta\, {\rm ch\,}\eta\,
   e^{-a(r)\,{\rm ch\,}\eta}
 \label{9}
 \end{equation}
with
 \begin{equation}
 a(r) = {m_{\perp}\over T}\,{{\rm ch\,}}{\eta}_t(r)\, ,
 \label{11}
 \end{equation}
the expressions (\ref{4}) for the HBT radii can be written as sums of
terms of the general form
 \begin{equation}
 {\langle{f(x,y)\,G^{mn}(r)}\rangle}_*
   \equiv {\int dx\, dy\, f(x,y)\, G^{mn}(r)\, F(x,y)
       \over
       \int dx dy \, F(x,y)}\, ,
 \label{10}
 \end{equation}
 where $f(x,y)$ is a simple polynomial in $x$ or $y$ and
 \begin{equation}
 F(x,y) =
   K_1(a(r))\, e^{K_\perp (x/r)\, {\rm sh\,}\eta_t(r)/T}\, e^{-r^2/(2 R^2)}
 \label{10a}
 \end{equation}
is the transverse weight function for the $\langle \dots \rangle_*$
average in (\ref{10}). The functions $G^{mn}$ needed in (\ref{4}) are
given explicitly by
 \begin{eqnarray}
 G^{01}(r) &=& {1\over a} + {K_0(a) \over K_1(a)} \, ,
 {\nonumber}\\
   G^{02}(r) &=& 1 + {K_2(a) \over a\, K_1(a)}  \, ,
 {\nonumber}\\
   G^{20}(r) &=& {K_2(a) \over a\, K_1(a)} \, ,
 \label{12}
 \end{eqnarray}
with $a=a(r)$ from (\ref{11}). With these definitions the HBT radii
(\ref{4}) take the form
 \begin{eqnarray}
R_l^2 &=& \tau_0^2\, \left\langle G^{20}(r) \right\rangle_*\, ,
 {\nonumber} \\
   R_s^2 &=& \left\langle y^2 \right\rangle_*\, ,
 {\nonumber} \\
   R_o^2 &=& \left\langle x^2 \right\rangle_*
           - \left\langle x \right\rangle_*^2
 {\nonumber} \\
         &-& 2 \beta_\perp \tau_0
            \left( \left\langle x\, G^{01}(r) \right\rangle_*
                 - \left\langle x \right\rangle_*
                   \left\langle G^{01}(r) \right\rangle_* \right)
 {\nonumber} \\
         &+& \beta_\perp^2 \tau_0^2
             \left(\left\langle G^{02}(r)\right\rangle_*
                 - \left\langle G^{01}(r)\right\rangle_*^2 \right) \, ,
 {\nonumber} \\
   R^2_{lo} &=& 0\, .
 \label{13}
 \end{eqnarray}
For the model (\ref{6}) these expressions are exact.

In some previous papers \cite{S95,AS95,MS88} the $\eta$-integration
was not done exactly, but by saddle point integration around $\bar\eta
=0$. To this end one expands the exponent of (\ref{8}) as ${\rm ch\,} \eta
\simeq 1 + \eta^2/2$, and replaces all ${\rm ch\,}\eta$ prefactors by
their saddle point value ${\rm ch}\,\bar\eta =1$. This amounts to
replacing the modified Bessel functions by their leading term for
large $a$, $K_\nu(a) \simeq \sqrt{\pi/(2a)}\, e^{-a}$. It is easy to
see that in this approximation the $\tau_0$-dependent terms in
Eq.~(\ref{13}) for $R_o^2$ vanish exactly. One would thus conclude
that for sharp freeze-out at $\tau=\tau_0$ there is no influence of
the source lifetime on $R_o^2$ and on the difference $R_o^2-R_s^2$.

The exact expressions (\ref{13}) show, however, that this is not true:
$R_o^2-R_s^2$ is sensitive to the time structure of the source. The
physical mechanism behind this is quite interesting: since particles
from different points on the freeze-out surface contribute to the
correlation function as long as they are separated in $z$-direction by
less than the longitudinal region of homogeneity $R_l$, the
correlation function indeed probes a finite range of coordinate times
$t$ along the proper-time hyperbola $\tau{=}\tau_0$: $\Delta
t \simeq \sqrt{\tau_0^2{+}R_l^2/4} - \tau_0$ (assuming that only half
of the longitudinal region of homogeneity contributes effectively).
This shows up in the HBT radius for the out-direction as a
contribution from a finite duration of the emission process and is
formally reflected by the last two lines in the expression (\ref{13})
for $R_o^2$. We will see the practical consequences of this in
Sections \ref{sec5} and \ref{sec6}.

\subsection{Transverse saddle point integration}\label{sec3B}

Since in most realistic situations we can assume $a(r) \geq {m_\perp
\over T}>1$ (low-$K_\perp$ photons from a high-temperature source
being the exception), it makes sense to use in the transverse
integrals the asymptotic expansion of the Bessel functions for large
arguments:
 \begin{equation}
 K_{\nu}(a) = \sqrt{\pi\over 2a} e^{-a}
   {\left({ \sum_{n=0}^{p-1}
   {1\over (2a)^n} {{\Gamma}(\nu + n + {1\over 2})\over
   n! {\Gamma}(\nu - n + {1\over 2})} + R(\nu,p,a)}\right)}\, .
 \label{14}
 \end{equation}
Clearly, how many orders $p$ must be taken into account to obtain a
good approximation for (\ref{10}) depends on both the order $\nu$
and the lower limit for $a$, as reflected by the upper bound
\cite{AS65}
 \begin{equation}
   \bigl|R(\nu,p,a)\bigr| <  {{\Gamma}(\nu + p + {1\over 2})\over
   (2a)^p p! |{\Gamma}(\nu - p + {1\over 2})|}\, , \qquad
   p \geq \nu - {1\over 2}\, ,
 \label{15}
 \end{equation}
for the remainder term in Eq.~(\ref{14}).

Substituting the expansion (\ref{14}) back into (\ref{13}), we are
left with sums of integrands of the type $f(x,y)\, G_n({\bf r})$ where
 \begin{equation}
   G_n({\bf r}) = ({\rm ch\,}\eta_t)^{-n-{1\over 2}}\,
   e^{-(m_\perp{\rm ch\,}\eta_t - K_\perp (x/r)\, {\rm sh\,}\eta_t)/T}\,
   e^{-r^2/(2 R^2)}\, .
 \label{16}
 \end{equation}
The integer $n$ labels the order in the expansion (\ref{14}). The
resulting transverse integrals can be done by a saddle point
approximation of $G_n({\bf r})$. To obtain accurate results it is,
however, important to find a good approximation for the exact saddle
point $\bar x_n$ of $G_n({\bf r})$ which cannot be obtained
analytically except for simple limiting cases \cite{S95}. In previous
studies \cite{CSH95,S95,AS95} a Gaussian approximation of the first
exponential factor in (\ref{16}) around its own saddle point was used;
the resulting product of two Gaussian factors was integrated out
analytically, without however including the effects from the $({\rm
ch\,}\eta_t)^{-\left(n+{1 \over 2} \right)}$ prefactor in (\ref{16}).
Unfortunately, we found that in most cases the resulting analytical
expressions for the HBT radii do not exhibit the correct $m_\perp$
dependence of the exact results obtained by evaluating the expressions
(\ref{13}) numerically. In Appendix~\ref{appa} we derive an analytical
approximation scheme for obtaining a better estimate of the true
saddle point which yields much more accurate results. We derive there
the following Gaussian approximation for the transverse weight
function $G_n({\bf r})$:
 \begin{equation}
 G_n({\bf r}) \approx C_n\,\exp\left[
            - {y^2 \over 2\, R_s^2(n) }
            - {(x - \bar x_n)^2  \over 2\ R_o^2(n) } \right] \, ,
 \label{17}
 \end{equation}
with effective Gaussian size parameters
 \begin{eqnarray}
 {1\over R_s^2(n)}
     &=&  {1\over R^2}
      + {1\over {\lambda}_s^2(\bar x_n)}\, ,
 {\nonumber} \\
   {1\over R_o^2(n)}
     &=& {1\over R^2}
      + {1\over {\lambda}_o^2(\bar x_n)}\, .
 \label{18}
 \end{eqnarray}
Here $\bar x_n$ is the $x$-component of the transverse saddle point
$\bar {\bf r}_n$ which should be determined from the condition
 \begin{equation}
 \left( \left(n + {1\over 2} + {m_\perp\over T} \right) {\rm sh\,} \bar \eta_n
         - {K_\perp\over T} {\rm ch\,} \bar \eta_n \right)
   {\bar \eta'_n \over \bar x_n }
   = - {1 \over R^2}\, .
 \label{18a}
 \end{equation}
Barred quantities with an index $n$ denote values at the saddle point,
and primes indicate $x$-derivatives. Since this equation cannot be
solved analytically for $\bar x_n$, one replaces in Eqs.~(\ref{18})
the true saddle point $\bar x_n$ by an analytically obtained
approximate value $\tilde x_n$ and calculates the `lengths of
homogeneity' $\lambda_s^2( \bar{x}_n )$ and $\lambda_o^2( \bar{x}_n )$
by evaluating the expressions (\ref{A6},\ref{A6a}) at that point. The
correct saddle point $\bar x_n$ is then related to the approximate
value $\tilde x_n$ by Eqs.~(\ref{A7},\ref{A9}), while the constant
$C_n$ is determined by Eq.~(\ref{A11}). A good choice for the
approximate saddle point $\tilde x_n$ is also derived in
Appendix~\ref{appa} and given analytically by Eqs.~(\ref{A12}),
(\ref{A4}) and (\ref{A6c}).

Please note that according to Eqs.~(\ref{A6},\ref{A6a}) the
homogeneity lengths in the `side' and `out' directions, $\lambda_s$
and $\lambda_o$, are in general different; this was also recently
found by Akkelin and Sinyukov \cite{AS95}. Only for a linear
transverse rapidity profile with a small slope (the nonrelativistic
transverse expansion scenario studied in Refs.~\cite{CSH95,CNH95}),
where the transverse saddle point can be taken as $\bar x_n \approx
0$, do the two expressions agree. We will see in Section~\ref{sec5}
that in some situations this difference can be drastic, causing
sizeable differences between $R_o$ and $R_s$ which have nothing to do
with a finite duration of the emission process.

Using the expansion (\ref{14}), the HBT radii (\ref{13}) can be
rewritten in terms of integrals over the auxiliary functions $G_n({\bf
r})$, Eqs.~(\ref{16}/\ref{17}). Both the numerator and denominator of
(\ref{10}) reduce to integrals of the type
 \begin{equation}
 \sum_{n=0}^p c_n {\left({T\over m_{\perp}}\right)}^{n+ {1\over 2}}
   \int dx\, dy \, f(x,y)\, G_n({\bf r})\, ,
 \label{19}
 \end{equation}
with certain coefficients $c_n$ which depend on the particular HBT
radius under consideration (see Appendix~\ref{appb}). For the
polynomials $f(x,y)$ occuring in (\ref{13}) the necessary integrals
are easily evaluated in terms of the size parameters (\ref{18}):
 \begin{eqnarray}
 \int dx\, dy\, G_n({\bf r})
   &=& 2\pi C_n\, R_o(n)\, R_s(n)\, ,
 {\nonumber} \\
   \int dx\, dy\, y^2\, G_n({\bf r})
   &=& 2\pi C_n\, R_o(n)\, R_s^3(n)\, ,
 {\nonumber} \\
   \int dx\, dy\, x\, G_n({\bf r})
   &=& 2\pi C_n\, R_o(n)\, R_s(n)\, \bar x_n \, ,
 {\nonumber} \\
   \int dx\, dy\, x^2\, G_n({\bf r})
   &=& 2\pi C_n\, R_o(n)\, R_s(n)\,
            \Bigl( R_o^2(n) + \bar x_n^2 \Bigr)\, .
 \label{20}
 \end{eqnarray}

Before combining these results into analytical expressions for the HBT
radii (see Appendix~\ref{appb} and Section~\ref{sec3C}), let us
shortly comment on the validity of the transverse saddle point
approximation introduced in this subsection. We have checked three
possible sources of deviations of the approximate analytical HBT radii
from the results obtained by exact numerical integration of the
expressions (\ref{13}):

\begin{itemize}

\item[(i)] The expansion (\ref{14}) of the Bessel functions
$K_{\nu}(a)$ is asymptotic, but not convergent. In our case this is
not a problem since we found that in all cases where the approximation
scheme worked (see below), convergence was reached at low orders $p
\leq 3$, cf. Section \ref{sec5A}.

\item[(ii)] The factor $({\rm ch\,}\eta_t)^{-\left(n+{1\over 2}\right)}$ in
$G_n({\bf r})$ was substituted by $\exp\left[ -\left(n+ {1 \over 2}
\right) \left({{\rm ch\,}\eta_t - 1} \right)\right]$, cf.
Appendix~\ref{appa}.  Numerical checks showed that in all the cases
considered in Section~\ref{sec5} the errors introduced by this
approximation are negligible.

\item[(iii)] The crucial approximation in our method turns out to be
the substitution of the transverse weight function $G_n({\bf r})$ (in
the form (\ref{A2})) by the Gaussian (\ref{17}). This is good only as
long as the quadratic term in the expansion of $\ln G_n({\bf r})$
around $\bar x_n$ dominates the behaviour of the integrand. As we will
discuss in Section~\ref{sec5B}, this places a strong mathematical
constraint on the form of the transversal flow $\eta_t(r)$.

\end{itemize}

\subsection{Expressions for the HBT radii}\label{sec3C}

It is now only a matter of patience to combine Eqs.~(\ref{10}),
(\ref{12}), (\ref{14}), (\ref{19}), and (\ref{20}) to obtain
analytical expressions for all the HBT radii (\ref{13}) in terms of
the two effective Gaussian size parameters (\ref{18}). This is done in
Appendix~\ref{appb}. Since the comparison in Section~\ref{sec5} with
numerical results will show that we typically must keep several terms
from the expansion of the Bessel functions, these analytical results
are rather complicated and don't lend themselves easily to an
intuitive interpretation. In particular, different terms of the
expansion generate different types of $m_\perp$ dependences, and there
is no simple $m_\perp$ scaling law for the HBT radii.

It is nevertheless instructive to compare our analytical results with
simpler ones previously proposed by several other authors
\cite{CL94,CSH95,S95,AS95,MS88,HB95}. Those results were attractive
because they implied simple scaling laws for the HBT radius parameters
as a function of the average transverse momentum $K_\perp$ of the
boson pair (resp. of its transverse mass $m_\perp$), and for this
reason they are still popular with experimentalists (see, e.g.,
\cite{F92,NA35,NA44}). We will discuss here which limits have to be
taken in order to recover these results from our expressions, and what
these limits imply.

Let us begin with the longitudinal HBT radius $R_l$. Historically
the first expression was derived by Makhlin and Sinyukov in a model
without transverse flow, via saddle point approximation around $\bar
\eta=\bar{\bf r}=0$. They found
 \begin{equation}
   R_l^2 = {\tau}_0^2 {T\over m_{\perp}}
   \qquad\hbox{[{\it Makhlin and Sinyukov} \cite{MS88}]} \, ,
 \label{21a}
 \end{equation}
and this has been used by experimentalists to fit their data
\cite{F92,NA35,NA44}. For boson emitting sources with a
non-relativistic linear transverse flow $v (r / R)$, $v\ll c$, a next
order correction to (\ref{21a}) was calculated by Chapman {\it et al.}
who found in the boost-invariant limit\footnote{
 For the sake of completeness, we mention another generalization of
the Makhlin-Sinyukov formula,
 $$
  R_l^2 = \tau_0^2 {T\over m_\perp} \,
  {1\over {1 + {T\over m_{\perp} (\Delta\eta)^2}}}
  \qquad\hbox{[{\it Cs\"org\H{o} and L\o rstad} \cite{CL94}]} \, ,
 $$
which was found by saddle point approximation for a source with a
finite longitudinal extension $\Delta \eta$ which breaks the
boost-invariance. In the limit $\Delta\eta \to \infty$, this
expression reduces to (\ref{21a}). The results of Ref.~\cite{CSH95}
generalize it by non-leading corrections to the saddle point
approximation. Our investigation here is restricted to boost-invariant
models.}
($\Delta\eta \to \infty$ in their notation)
 \begin{equation}
   R_l^2 = \tau_0^2 {T\over m_\perp} \,
   \left( 1 +
   \left( {1 \over 2} + {1\over {1 + {m_\perp \over T} v^2}} \right)
          {T\over m_\perp} \right)
   \qquad\hbox{[{\it Chapman, Scotto and Heinz} \cite{CSH95}]} \, .
 \label{21b}
 \end{equation}
In that paper the transverse shift of the saddle point away from $\bar
{\bf r}=0$ was neglected, which is justified for small transverse flow
velocities. An analytical expression without any approximation was
recently derived for the case of vanishing transverse flow:
 \begin{equation}
 R_l^2 = \tau_0^2 {T\over m_\perp} \,
   { K_2({m_\perp \over T}) \over K_1({m_\perp \over T}) }
   \qquad\hbox{[{\it Herrmann and Bertsch} \cite{HB95}]}\, .
 \label{21c}
 \end{equation}
In the limit $v\to 0$, (\ref{21b}) reduces to the first two terms from
an expansion of (\ref{21c}) for $m_\perp \gg T$. Our expression
(\ref{13}) for the longitudinal radius in the case of arbitrary
transverse flow reads (cf. Appendix \ref{appb})
 \begin{equation}
   R_l^2 = \tau_0^2
   \left \langle {1\over a(r)}\,
   {K_2({1\over a(r)}) \over K_1({1\over a(r)}) }
   \right \rangle_{\!*}
   = \tau_0^2 \, {\sum_{n=0}^p \tilde{c}_n F_n \over {\sum_{n=0}^p c_n F_n}}
  \qquad\hbox{[{\it our general expression}]}\, .
  \label{21d}
\end{equation}
{}From this result Eq. (\ref{21c}) follows immediately in the limit of
vanishing transverse flow, $\lim_{\eta_t=0} a(r) = {m_\perp \over T}$, and
Eq.~(\ref{21a}) is the leading term of an expansion of (\ref{21c})
for large ${m_\perp \over T}$. It was pointed out before \cite{HB95} that
keeping only this lowest order term is an insufficient approximation for
realistic values of $T$ and $m_\perp$, and that therefore (\ref{21a})
should not be used to extract the freeze-out time $\tau_0$ from data.

In the presence of transverse flow the $m_\perp$-dependence of $R_l$
is much more complicated than either (\ref{21a}) or (\ref{21c}). The
approximate result (\ref{21b}) for a weak linear transverse flow
cannot, in this form, be derived from (\ref{21d}); one can show that
for $v^2 \ll {T\over m_\perp} \ll 1$ the leading term $\sim v^2$
agrees in both expressions, but differences occur in higher orders.
These arise from the transverse shift of the saddle point $\bar {\bf
r}$ which was neglected in Ref.~\cite{CSH95}, but must be taken into
account for stronger transverse flows. This leads to a significant
further modification of the $m_\perp$-dependence which will be studied
in detail in the following Section.

Let us now turn to the ``side"- and ``out"-radii. Here we can again
compare\footnote{
 The results of Akkelin and Sinyukov \cite{AS95} for systems with
strong transverse flow, obtained by saddle point approximation, are
difficult to compare with since the authors have for mathematical
convenience modified the geometrical part of the emission function
(\ref{8}).}
 with the expressions obtained by Chapman {\it et al.} \cite{CSH95}
for a weak linear transverse flow, expanding around $\bar {\bf r}=0$.
(Earlier results by Cs\"org\H o and L{\o}rstad \cite{CL94} can be
obtained from those of Ref.~\cite{CSH95} by keeping only the leading
terms of this expansion.) In the limit of a boost-invariant source
($\Delta\eta \to \infty$ in their notation) and sudden freeze-out at
proper time $\tau_0$ ($\Delta \tau \to 0$) they find in the LCMS frame
($\beta_L=0$)
 \begin{equation}
 \left.
   \begin{array}{l}
    {\displaystyle{
    R_o^2 = {R^2 \over 1 + {m_\perp \over T} v^2}
    + \textstyle{1\over 2} {\left({T\over m_{\perp}}\right)}^2
    {\beta}_{\perp}^2 {\tau}_0^2 \, ,
    }}
    \\[1ex]
    {\displaystyle{
    R_s^2 = {R^2 \over 1 + {m_\perp \over T} v^2} \, .
    }}
   \end{array}\right\}
  \qquad\hbox{[{\it Chapman, Scotto and Heinz} \cite{CSH95}]}
  \label{22}
 \end{equation}
Let us compare this with the lowest order ($n=0$) contributions from
our approximation scheme (see Appendix \ref{appb}):
 \begin{equation}
 \left.
   \begin{array}{l}
    {\displaystyle{
    {1\over R_s^2} = {1\over R_s^2(0)}
    = {1\over R^2} + {1\over {\lambda}_s^2(\bar{x}_0)}\, ,
    }}
    \\[1ex]
    {\displaystyle{
    {1\over R_o^2} = {1\over R_o^2(0)}
    = {1\over R^2} + {1\over {\lambda}_o^2(\bar{x}_0)}\, .
    }}
   \end{array}\right\}
   \qquad\hbox{[{\it our lowest order contribution}] }
 \label{23}
 \end{equation}
For a weak linear transverse flow $\eta = \eta_{\rm f} (r/R)$,
$\eta_{\rm f} \ll 1$, one may approximate the transverse saddle point
by $\bar x_0{=}0$. Then with a little algebra
Eqs.~(\ref{A6}/\ref{A6a}) can be shown to yield ${1\over \lambda_o^2
(0)} = {1\over \lambda_s^2(0)} = A_0 {\eta_{\rm f}^2 \over R^2}$.
Inserting this into (\ref{23}) one recovers the first term on the
right hand sides of Eqs.~(\ref{22}) except that in the denominator
${m_\perp \over T}$ is replaced by $A_0 = {m_\perp \over T} + {1 \over
2}$. The reason for this slight discrepancy is that in
Ref.~\cite{CSH95} (as in all other previous discussions of the emission
function (\ref{8})) the contribution of the factor $({\rm ch\,}\eta_t)^{-n-
{1\over 2}}$ in (\ref{16}) to the lengths of homogeneity was
overlooked.

These first terms in Eqs.~(\ref{22}) arise from the ``geometric"
parts, ${\langle{x^2}\rangle} - {\langle{x}\rangle}^2$ and
${\langle{y^2}\rangle}$, respectively, in the model-independent
expressions (\ref{4}), which for weak linear transverse flow coincide
to lowest order of our approximation scheme. The second term in the
expression (\ref{22}) for $R_o$ is due to the finite duration of
particle emission by the source (\ref{8}) and comes from the term
$\beta_\perp^2 \left( \langle t^2 \rangle - \langle t \rangle^2
\right)$ in (\ref{4}). It is identical with the lowest non-vanishing
contribution from Eq.~(\ref{B6}). Thus the authors of \cite{CSH95}
have correctly identified the lowest order contributions to $ \langle
x^2 \rangle - \langle x \rangle^2$, $\langle y^2 \rangle$, and
$\beta_\perp^2 \left( \langle t^2 \rangle - \langle t \rangle^2
\right)$. (The term $-2 \beta_\perp \left( \langle xt \rangle -
\langle x \rangle \langle t \rangle \right)$ in $R_o^2$ is much
smaller and according to (\ref{B5}) can only be seen if the effect of
the transverse flow on the saddle point $\bar x_i$ is taken into
account.) In the following Section we will show, however, that for
realistic values of the transverse flow velocity these lowest order
contributions are unreliable, and a correct description of the
$K_\perp$-dependence of the HBT radii requires the inclusion of higher
order corrections from our general expressions in Appendix \ref{appb}.

We already noted that for transverse flows which are not weak or have
a non-linear dependence on $r$, the two transverse homogeneity lengths
${\lambda}_o$ and ${\lambda}_s$ are no longer equal (see also
\cite{AS95}). In that case it is no longer guaranteed that the finite
duration of particle emission yields the dominant contribution to the
difference $R_o^2-R_s^2$ as in models with weak or vanishing
transverse flow \cite{CSH95,CNH95,CP91}. Since transverse flow is a
generic feature of the sources generated in heavy-ion collisions
\cite{SH94}, its effect must to be taken into account when trying to
interpret measured differences between $R_o$ and $R_s$ in terms of the
emission time.

\section{Numerical evaluation of the correlator}\label{sec4}

The $\eta$-integration for the full 2-particle correlation function
$C({\bf K},{\bf q})$ as given by Eq.~(\ref{2}) can be done
analytically by the same methods as used in Section \ref{sec3A}. The
resulting 2-dimensional transverse Fourier integrals will then be
evaluated numerically. The half widths of the numerically computed
exact correlator can then be compared with the analytical expressions
for the HBT radii from the previous Section, allowing us to check our
approximations.

The $\eta$-integral for the denominator of Eq.~(\ref{2}) was already
evaluated in Section~\ref{sec3A}:
 \begin{equation}
   \int d^4x\, S(x,K) = {2 \tau_0 m_\perp \over (2\pi)^3}
   \int dx\,dy\, K_1(a(r))\, e^{ K_\perp (x/r)\,{\rm sh\,}{\eta}_t/T}
   \, e^{-r^2/(2R^2)}
   \, .
 \label{24}
 \end{equation}
For the numerator let us first study the limit $q_l =0$. Writing
$t{=}\tau\,{\rm sh\,}\eta$, $z{=}\tau\,{\rm ch\,}\eta$ and using
$q^0{=}\beta_\perp q_o$ (which is true in the LCMS frame where
$\beta_L=0$), we find
 \begin{eqnarray}
 && \int d^4x\, S(x,K)\, e^{i(\beta_\perp q_o t - q_o x - q_s y)}
 {\nonumber} \\
  && = {2 \tau_0 m_\perp \over (2\pi)^3}
     \int dx\, dy\, e^{-i(q_o x + q_s y)}\,
     K_1\Bigl(a(r){-}i \tau_0 \beta_\perp q_o\Bigr) \,
     e^{K_\perp (x/r)\,{\rm sh\,}\eta_t(r)/T}\,
     e^{-r^2/(2R^2)} \, .
 \label{25}
 \end{eqnarray}
The $q_l$-dependence of $C({\bf K},{\bf q})$ can be obtained by
using
 \begin{equation}
 \int_{-\infty}^\infty d\eta\, {\rm ch\,}\eta\  e^{-a\,{\rm ch\,}
 \eta - i\,b\,{\rm sh\,}\eta}
   = {2\,a\over \sqrt{a^2 + b^2}} \,
   K_1\left( \sqrt{a^2 + b^2} \right)
 \label{25a}
 \end{equation}
for real values of $a$ and $b$. This leads to
 \begin{eqnarray}
  && \int d^4x\, S(x,K) \, e^{-iq_lz}
 {\nonumber} \\
  && = {2 \tau_0 m_\perp \over (2\pi)^3}
   \int dx\, dy\,
   {a(r) \over \sqrt{a(r)^2 + q_l^2 \tau_0^2} }\,
   K_1\left( \sqrt{a(r)^2 + q_l^2 \tau_0^2} \right)\,
   e^{ K_\perp (x/r)\, {\rm sh\,}\eta_t(r)/T}\,
   e^{-r^2/(2R^2)} \, .
 \label{26}
 \end{eqnarray}
With these (exact) expressions the correlation functions $C({\bf K},
q_o, q_s, q_l{=}0)$ and $C({\bf K},q_o{=}q_s{=}0,q_l)$ reduce to a
ratio of 2-dimensional integrals. These have been evaluated
numerically. From the numerical result for the correlator we determine
new HBT radii $R_i^c({\bf K})$ by requiring that the function
 \begin{equation}
   C({\bf K},q_i) = 1 + e^{-R_i^c({\bf K})^2 q_i^2}\, , \qquad
   i=o,\, l,\, s,\,
 \label{27}
 \end{equation}
reproduces exactly the half point, $C({\bf K}, {\bf q}) = 1 +
{1\over 2}$, of the numerically computed correlator (the other two
components $q_{i\ne j}$ are set equal to zero). In the following
Section these values $R_i^c({\bf K})$ will be compared with HBT radii
obtained from the expressions (\ref{13}) by either numerical or
analytical evaluation of the transverse integrals $\langle \dots
\rangle_*$.

\section{Results}\label{sec5}

In this section we give a detailed and quantitative analysis of the
HBT radii for the source function (\ref{6}/\ref{7}), both for a linear
and a quadratic transverse flow rapidity profile $\eta_t(r)$. We will
compare the analytic methods developed in Section~\ref{sec3} with
exact numerical results. We will show that for the linear transverse
flow the analytical approximation scheme works very well, but only if
higher order contributions (non-leading terms in the Bessel function
expansion and an improved value for the transverse saddle point) are
properly taken into account. For the quadratic flow the saddle point
methods from Section~\ref{sec3} are found to fail. For a weak
quadratic transverse flow we also find an unexpected effect: in the
$x{-}y$-plane the emission region {\em increases} with rising
transverse momentum of the pair. Although this case may be somewhat
pathological, it provides a specific counter example to the folklore
\cite{P86,CL94,CSH95} that transverse collective flow leads to a
reduction of the transverse HBT radii at finite transverse pair
momentum.  Another example with a different velocity profile, but
similar behaviour for $R_o$ was found in Ref.~\cite{AS95}.

\subsection{Linear transverse flow rapidity profile}\label{sec5A}

All the sources studied in this Section have a transverse geometric
(Gaussian) radius $R=3$ fm and freeze out along a hyperbola of
constant longitudinal proper time $\tau_0 = 3$ fm/$c$ at temperature
$T=150$ MeV. The longitudinal flow is always boost-invariant, and we
study the sensitivity to the transverse flow.

Let us begin with a linear transverse flow rapidity profile:
 \begin{equation}
  \eta_t(r) = \eta_{\rm f}\, {r \over R} \, .
 \label{27a}
 \end{equation}
For this case the transverse momentum dependence of the HBT radii
$R_s$, $R_o$, and $R_l$ is shown in Fig.~\ref{F1}. In the figure we
compare various approximations: The solid lines are obtained
by numerical evaluation of the expectation values in the
model-independent expressions (\ref{4}) in the form (\ref{13}) with
our source (\ref{6}). The long-dashed lines show the radii $R_i^c$
from Eq.~(\ref{27}) which reproduce the half width of the exact
correlation function in direction $i$. The short-dashed lines
represent our analytical results (\ref{B2}) - (\ref{B6}), evaluated
with the effective size parameters (\ref{18}) obtained in
Appendix~\ref{appa} and including all terms up to order $p=3$ from the
expansion of the Bessel functions. The dash-dotted lines show the same
analytic expressions but truncating the expansion (\ref{14}) at lowest
order $p=0$.

Let us summarize the most important features of Fig.~\ref{F1}:

\begin{itemize}

\item[1.] For $R_s$ our analytical expression (\ref{B2}) approximates very
accurately, even for $p=0$, the exact value obtained numerically from
Eq.~(\ref{13}). Thus for $R_s$ the leading order expression
 \begin{equation}
  {1\over R_s^2} = {1\over R^2}
                  + {1\over \lambda_s^2(\bar{x}_0)}
 \label{28}
 \end{equation}
can be used with excellent accuracy. However, it is necessary that the
homogeneity length $\lambda_s$ is evaluated sufficiently closely to
the exact transverse saddle point $\bar x_0$ as described in
Appendix~\ref{appa}. Our studies showed that, if used with the lowest
order estimate (\ref{A6b}), Eq.~(\ref{28}) develops for large
transverse flow rapidities $\eta_{\rm f} > 0.3$ a much stronger
$K_\perp$-dependence than the exact ``side"-radius. Clearly this
renders an analytical determination of the transverse flow velocity
from the $K_\perp$-dependence of $R_s$ a somewhat subtle issue, to say
the least.

\item[] The exact numerical value for $R_s$ from Eq.~(\ref{13}) also
coincides very accurately with $R_s^c$ from (\ref{27}). For
the source (\ref{6}) our model-independent expressions (\ref{4}) thus
correctly reproduce not only the curvature of the correlation function
at small values of $q_s$ (which is difficult to access experimentally), but
also its half width. This should not be too surprising after one
checks that in the ``side"-direction the source (\ref{6}) has a rather
Gaussian shape for all relevant values of $K_\perp$.

\item[2.] In the ``out"-direction the situation is more complicated:
there the leading term from the Bessel function expansion is seen to
yield qualitatively wrong results. After including higher order terms
up to $p{=}3$, the agreement with the exact numerical results improves
dramatically, although some deviations remain at large $K_\perp$ for
strong transverse flows. This remaining discrepancy can be traced back
to the fact that we use for the transverse integration only an
approximate saddle point. For the ``naive" saddle point $x_n^d$ the
disagreement is much worse. This shows that for the ``out"-direction
and strong transverse flow the usefulness of the analytical approach
relies crucially on an accurate approximation for the transverse
saddle point.

\item[] It is important to note that the lowest order approximation
(dash-dotted line) completely misses the rise of $R_o$ at small values
of $K_\perp$. This is a lifetime effect and arises from the time
variance given in (\ref{B3}). The rise was also noted previously in
correlation functions based on numerical hydrodynamic calculations
\cite{Marb}, but without a detailed theoretical analysis of its
origin. As seen from Eq.~(\ref{B9}), the time variance only begins to
receive contributions at third order of the Bessel function expansion,
but the coefficients of these higher order terms are larger than
unity. The physics of this effect has been explained at the end of
Section~\ref{sec3A}. For weak flow ($\eta_{\rm f} \leq 0.3$) it yields
the dominant contribution to the difference $R_o^2-R_s^2$ between the
``out" and ``side" radii.

\item[] For strong transverse flow, that difference remains
appreciable even if the contribution from the time variance is
neglected; it is then due to the difference between the homogeneity
lengths in the two directions, see Eqs.~(\ref{A6}/\ref{A6a}). This
difference can be rather accurately estimated by comparing the
dash-dotted lines in Fig.~\ref{F1}a (our lowest order analytical
results for $R_o$ which give essentially $\sqrt{ \langle x^2 \rangle -
\langle x \rangle^2 }$) with Fig.~\ref{F1}b. For $\eta_{\rm f} = 0.9$
one finds, for example, that in the region $K_\perp > 500 $ MeV/$c$
the expressions $\langle y^2 \rangle$ and $\langle x^2 \rangle -
\langle x \rangle^2$ differ by more than 50\%.

\item[] The agreement between the exact HBT radius $R_o$ from the
model-independent expression (\ref{4}) and the half widths described
by $R_o^c$ is excellent for all values of $\eta_{\rm f}$ and
$K_\perp$. The reason is again the rather Gaussian shape of the source
function also in the ``out"-direction.

\item[3.] For the longitudinal radius $R_l$, the leading term from the
Bessel function expansion is insufficient again (this is even true for
vanishing transverse flow, see \cite{HB95}), but excellent agreement
with the exact value from the model-independent expression (\ref{4})
is reached for $p = 3$, for all values of $K_\perp$. For small
transverse momenta, however, one sees a disagreement with the half
width of the exact correlator as given by the parameter $R_l^c
(K_\perp)$. On the other hand we checked numerically that $R_l^2
(K_\perp)$ from Eq.~(\ref{4}) correctly reproduces the curvature of
$C(K_{\perp},q_l)$ at $q_l =0$, as it should. In this case the
disagreement arises from non-Gaussian features of our source (\ref{6})
in the longitudinal ($\eta$) direction: for small $K_\perp$ the source
decreases at large values of $\eta$ much more steeply than a Gaussian.
However, the discrepancy is small, and in spite of these non-Gaussian
features of the source the model-independent radii (\ref{4}),
evaluated with the full source (\ref{6}), reproduce the half
width of the correlation function with an accuracy of a few percent.

\item[4.] All three HBT radii are strongly affected by the transverse
flow, even at $K_\perp=0$. The influence on $R_l$ is relatively weaker
than on $R_s$ and $R_o$. As transverse flow increases, the transverse
region of homogeneity decreases, and the HBT correlator receives
contributions from a smaller and smaller fraction of the total
source. At $K_\perp=0$ the ``out-" and ``side-" radii are equal,
$R_s(K_\perp{=}0){=}R_o(K_\perp{=}0)$; this must be true because for
$K_\perp=0$ there is nothing to distinguish between the two transverse
directions \cite{CSH95}.

\item[] An extraction of the geometric transverse radius $R$ from the
data requires to disentangle geometrical from dynamical effects. Based
on Fig.~\ref{F1} it appears to us that the most promising way to
achieve this is as follows: one estimates the amount of transverse
flow from the strength of the $K_\perp$-dependence of the
``side"-radius $R_s$. One then extrapolates $R_s(K_\perp)$ to
$K_\perp=0$ and uses the estimate for the transverse flow rapidity to
correct it for flow effects, thus extracting the geometric transverse
radius $R$.

\item[5.] From Fig.~\ref{F1}a it is clear that the extraction of the
duration of particle emission, which is related to the rate at which
$R_o(K_\perp)$ rises for small $K_\perp$, requires an accurate
measurement of the $K_\perp$-dependence of the ``out"-radius in the
domain $K_\perp < 100$ MeV/$c$. This is not easy, because for very
small $K_\perp$ the detection efficiency tends to decrease (in a TPC
the track density becomes high, in a single arm spectrometer one of
the two particles tends to miss the detector). It therefore helps to
know that $R_o$ agrees with $R_s$ at $K_\perp=0$; the latter can be
extracted much more easily by extrapolation, due to its smooth
$K_\perp$-dependence.

\end{itemize}

Fig. \ref{F1} shows that the leading terms from a saddle point
approximation produce quantitatively and qualitatively unreliable
results for the $m_\perp$-dependence of the HBT radii $R_l$ and $R_o$.
Quantitative attempts to extract values for the freeze-out time
$\tau_0$, the transverse flow $\eta_{\rm f}$ and the transverse
geometric size $R$ from HBT data must thus be based on a numerical
evaluation of the model-independent expressions (\ref{4}) rather than
on simple $m_\perp$-scaling laws extracted from insufficient
analytical approximations.

\subsection{Quadratic transverse flow rapidity profile}\label{sec5B}

A linear transverse flow rapidity profile is perhaps the simplest, but
not obviously the most realistic assumption. Dynamical studies
based on a numerical solution of the hydrodynamic equations
\cite{Ornik} produce transverse flow velocity profiles which for small
values of $r$ look parabolic and for large $r$ saturate at the light
velocity; the parabolic rise is zero initially, but builds up
gradually as the hydrodynamic pressure generates transverse flow. On
the other hand, when combining such hydrodynamic calculations
with a dynamic model for the freeze-out kinetics, taking into account
sequential transverse freeze-out of the matter from the dilute edge at
early times to the center at later times, the resulting flow profile
{\em along the freeze-out hypersurface} turns out \cite{UM95} to be
{\em linear} again for small $r$, reaching a maximum at intermediate
$r$ and dropping to zero again at the edge of the firetube (because
that part of the matter freezes out immediately due to its diluteness,
before transverse flow is able to build up). Numerical kinetic
simulations of pion production based on the RQMD code, finally,
produce a transverse flow profile at freeze-out \cite{Xu} which rises
quadratically at small $r$, reaches a maximum and slightly drops again
at large $r$.

The shape of the transverse flow profile is thus not at all clear.
Investigations of particle freeze-out based on hydrodynamic or kinetic
models, which result in source functions which are only numerically
known, and their influence on the shape of the HBT correlation
function should thus be supplemented by systematic studies of simple
model sources with various types of simple flow profiles. In this
subsection we therefore supplement the results from the previous one
by a discussion of quadratic transverse flow rapidity profiles:
 \begin{equation}
  \eta_t(r) = \eta_{\rm f}\, {r^2 \over R^2} \, .
 \label{28a}
 \end{equation}
The resulting transverse momentum dependence of the HBT radii
is shown in Fig.~\ref{F2}. The labelling and meaning of the various
curves is the same as in the previous subsection. However, the curves
corresponding to analytical approximations are missing; the reason is
that the scheme developed in Section \ref{sec3} and Appendix
\ref{appa} fails for a quadratic transverse flow rapidity profile.
That scheme relied on a Gaussian saddle point approximation in the
transverse direction; since for small $K_\perp$ the dominant
contribution from the transverse flow arises from the factor ${\rm ch\,}
\eta_t$ in the exponent which contributes only at order $r^4$, a
Gaussian saddle point approximation misses the dominant flow effects
completely. An improved analytical approximation scheme on the other
hand would be much more complicated.

Comparing Figs. \ref{F1} and \ref{F2} we see that qualitatively the
$K_\perp$-dependence of the HBT radii is similar for both types of
transverse flow profile. In all cases the model-independent
expressions (\ref{4}) yield HBT radii which give an excellent
representation of the half-width of the correlator. For a given flow
scale $\eta_{\rm f}$, the flow effects on the HBT radii are stronger
for the quadratic flow (\ref{28a}) than for the linear flow
(\ref{27a}). This indicates that emission from regions $r>R$ in the
source plays a significant role for the shape of the correlation
function.

The only unusual feature in Fig.~\ref{F2} is the rise of $R_o$ with
$K_\perp$ for a weak quadratic transverse flow with $\eta_{\rm f}=0.1$.
It can be traced to a rise of the variance in the ``out"-direction,
$\langle x^2 \rangle - \langle x \rangle^2$, with $K_\perp$.  This is
different from the generic decrease of the effective region of
homogeneity with increasing $K_\perp$ which is seen in all other
cases. It seems to be due to an accidental coincidence of parameters:
In Fig.~\ref{F3} we show a contour plot of the transverse weight
function (\ref{10a}); only for small quadratic transverse flow it gets
wider in the $x$ direction with increasing $K_\perp$. Although exotic,
this example should be kept in mind as a reminder that transverse flow
not always leads to a decrease of the effective source size.

\section{Summary and Conclusions}\label{sec6}

The aim of HBT interferometry with hadrons produced in heavy ion
collisions is to obtain information about the space-time structure of
the emitter. Unfortunately, a unique reconstruction of the emission
function $S(x,K)$ from the 2-particle correlation function $C({\bf q},
{\bf K})$ is impossible, since the frequency $q^0$ and the spatial
momenta ${\bf q}$ in the Fourier transformation relating the two
functions are not independent, as the final state hadrons are on
mass-shell. However, the class of possible models for the emission
function can be strongly restricted by a careful multidimensional HBT
analysis of the 2-body correlation function. In particular, a careful
analysis of the dependence of the HBT radii (which describe the width
of the correlator as a function of the relative momentum ${\bf q}$) on
the pair momentum ${\bf K}$ provides crucial information on the
correlations between emission point $x$ and particle momentum $K$ in
the source. Such $x{-}K$ correlations are generated, for example, by
the collective expansion of the sources generated in heavy ion
collisions. Recently heavy ion experiments \cite{F92,NA35,NA44} have
begun to provide first quantitative information on the
${\bf K}$-dependence of the HBT correlation functions. The hope to
obtain access to the collective behaviour in heavy ion collisions by
analyzing these results has refuelled the theoretical interest in the
${\bf K}$-dependence of the HBT radii.

In this paper we reanalyzed this issue with a combination of
numerical and analytical methods. For a given source model, the
numerical methods provide us with exact predictions for the transverse
momentum dependence of the HBT radii. They thus establish a reliable
link between various possible models for the emission function and the
experimental data for the correlation function. The analytical
approach provides, in the context of a specific source model studied
in this paper, a mathematical understanding of the numerical results
and a bridge between approximate analytical expressions published
previously and the exact numerical results obtained here, and thus
permits to test their reliability.

Our numerical studies showed that the model-independent expressions
(\ref{4}) for the HBT radii not only give the correct curvature of the
correlation function $C({\bf q}, {\bf K})$ for small values of ${\bf
q}$ (as was known before \cite{CSH95,HB95}), but also reproduce on
most cases the width of the correlator quite accurately. As long as
the measured correlation function has a roughly Gaussian shape in
${\bf q}$, these expressions provide an accurate and valuable link
between theory and experiment and permit to relate the
(${\bf K}$-dependent) width parameters of the correlation function to
the (${\bf K}$-dependent) space-time width of the original source
function. In this context it should be cautioned, however, that the
effects of resonance decays, which are numerically known to give the
correlation function a non-gaussian shape \cite{Marb,UM95}, remain to
be studied in more detail. Furthermore, this result means that {\em
only} the half-widths of the emission function in space-time can be
reconstructed from the width of the correlation function. Finer
details of the space-time structure of the emitter (like sharp edges
or holes in the center) do not affect the width of the correlator, but
can only be estimated by a very accurate study of the long-range decay
of the correlator (e.g. by looking for side maxima at larger values of
${\bf q}$).

The numerical study also enables us to give a detailed assessment of
the accuracy of previously published approximate analytical results
for the HBT radius parameters. This is of particular interest since
some of these results have gained great popularity because of their
simple scaling behaviour as a function of $m_\perp$. Here our results
cause disappointment: none of the so far suggested simple scaling laws
is quantitatively reliable, except for very limiting situations which
are not likely to be established in experiments. To make quantitative
use of the measured ${\bf K}$-dependence of the correlation function
and to obtain reasonably accurate estimates for the lifetime,
transverse geometric size and collective expansion rate of the source,
a full-blown numerical evaluation of the expressions (\ref{4}) is
required, at least. The simple scaling laws should not be used -- we
showed that they yield badly misleading results.

For the specific source studied in this paper the duration of particle
emission is very short, of order $\Delta t(K_\perp) \simeq \sqrt{
\tau_0^2 + R_l^2(K_\perp)/4 } - \tau_0$ ($\approx 0.9$ fm/$c$ at
$K_\perp=0$). This leads to a rise of $R_o$ at small $K_\perp$. The
rise stops at $K_\perp \simeq m_\pi$ (when $\beta_\perp$ approaches
unity), and beyond that point $R_o$ and the difference between $R_o-
R_s$ decreases again (because $\Delta t(K_\perp)$ decreases). The
experimental verification of the effect of a finite duration of
particle emission thus requires a high-statistics study of the
correlation function at small $K_\perp$. One might argue that
realistic models in which the sharp freeze-out along a proper time
hyperbola is replaced by continuous freeze-out over a longer time
period should lead to a larger effect which is more easily seen in the
data. However, in such models usually the source shrinks as the
particles evaporate, and particles emitted at later times come from
smaller regions near the center of the fireball; the net effect on the
difference $R_o - R_s$ seems to remain small and concentrated at small
$K_\perp < 100$ MeV/$c$ \cite{UM95}. In the light of these new
expectations the failure of the experiments so far
\cite{F92,NA35,NA44} to provide positive evidence for a finite period
of particle emission appears less puzzling. We no longer believe that
these data show that in heavy-ion collisions pion emission occurs ``in
a flash".


\acknowledgments

We would like to thank S. Chapman and U. Mayer for clarifying and
stimulating discussions. This work was supported in part by BMBF,
DFG, and GSI.


\appendix
\section{Determination of the saddle point}\label{appa}

Here we give the technical details of our saddle point approximation
of expression (\ref{16}). We begin by writing the prefactor as
$({\rm ch\,} \eta_t)^{-(n+{1\over 2})} = \exp[-(n+{1\over 2})
\ln({\rm ch\,}\eta_t)]$
and expanding the
logarithm around ${\rm ch\,} \eta_t = 1$:
 \begin{equation}
   ({\rm ch\,}\eta_t)^{-(n+{1\over 2})} = e^{-(n+{1\over 2})({\rm ch\,}
   \eta_t-1)}
   \left( 1 + {2n+1\over 16}\eta_t^4 + O(\eta_t^6) \right)\, .
 \label{A1}
 \end{equation}
We checked numerically that keeping only the first term in the
brackets is sufficient in the following. This allows us to combine the
exponent with the ${\rm ch\,} \eta_t$-term from the Boltzmann weight
factor and to rewrite $G_n({\bf r})$ in the form
 \begin{equation}
  G_n({\bf r}) \approx \exp \Bigl( n + {1\over 2} + d_n({\bf r})
                             - r^2/(2 R^2) \Bigr)\, ,
 \label{A2}
 \end{equation}where
 \begin{equation}
   d_n({\bf r}) = - A_n\, {\rm ch\,}\eta_t(r) + B\, {x\over r}\,
 {\rm sh\,}\eta_t(r)
 \label{A2a}
 \end{equation}
with
 \begin{equation}
 A_{n} = n + {1\over 2} + {m_\perp\over T} \, ,
   \qquad
   B = {K_\perp \over T} \, .
 \label{A3}
 \end{equation}
The saddle point ${\bf r}^d_n = (x^d_n, y^d_n)$ of the modified
dynamical term $\exp \bigl( d_n({\bf r}) \bigr)$ alone is found to
satisfy
 \begin{equation}
   y^d_n = 0\, ,  \qquad
   {\rm th\,} \eta^d_n \equiv
   {\rm th\,}\eta_t(x^d_n,0) =
   {K_\perp \over m_\perp + (n+{1\over 2}) T} = {B\over A_n}\, .
 \label{A4}
 \end{equation}
Omitting the term $\left(n+{1\over 2}\right) T$ in the denominator,
this reduces to the expression given in \cite{AS95} whose shortcomings
were discussed in section \ref{sec5A}. In this case the saddle point
is given by that point of the source at which the transverse flow
velocity ${\rm th\,}\eta_t$ agrees with the transverse velocity
$\beta_\perp{=}(K_\perp/m_\perp)$ of the pion pair. Numerically a
gaussian approximation around this point leads, however, to
unsatisfactory results. Including the effects of the $({\rm ch\,}
\eta_t)^{-(n+{1\over 2})}$ prefactor both in the saddle point and in
the gaussian curvature considerably improves the approximation. In
addition, the saddle point $\bar {\bf r}_n$ of the full function
$G_n({\bf r})$ is further shifted away from ${\bf r}^d_n$ by the
geometric factor $\exp \bigl(-r^2 / (2R^2) \bigr)$ in (\ref{A2}). Due
to the symmetry of the function $G_n({\bf r})$ this shift will be only
in $x$-direction, i.e. $\bar y_n$ will stay zero.

To estimate the position of the exact saddle point $\bar {\bf r}_n{=}(
\bar x_n, 0)$ we proceed as follows: we expand $\exp \bigl( d_n({\bf
r}) \bigr)$ to second order around an arbitrary point $\tilde
{\bf r}_n{=}(\tilde x_n,0)$. After combining the resulting gaussian with the
geometric factor $\exp \bigl(-r^2 / (2R^2) \bigr)$ we try to adjust
$\tilde x_n$ iteratively to obtain an analytical approximation to the
true saddle point $\bar x_n$ of the full weight function.

The quadratic expansion of the dynamical term reads
 \begin{eqnarray}
   d_n (\tilde{x}_n + {\delta}x,{\delta}y)
   &=& - \Bigl( A_n\, {\rm ch\,}\tilde\eta_n -
                B\, {\rm sh\,}\tilde\eta_n \Bigr)
       - \Bigl( A_n\, {\rm sh\,}\tilde\eta_n -
                B\, {\rm ch\,}\tilde\eta_n \Bigr)
         \tilde\eta'_n \, {\delta}x
 \nonumber \\
   & & - {{\delta}x^2 \over 2\lambda_o^2(\tilde{x}_n)}
       - {{\delta}y^2 \over 2\lambda_s^2(\tilde{x}_n)}\, ,
 \label{A5}
 \end{eqnarray}
where $\tilde\eta_n{\equiv}\eta_t(\tilde {\bf r}_n)$, $\tilde
\eta'_n{\equiv}(\partial \eta_t/\partial x)(\tilde x_n,0)$,
etc., and
 \begin{eqnarray}
   {1 \over \lambda_s^2(\tilde x_n)}
   &=& \Bigl( A_n\, {\rm sh\,}\tilde\eta_n -
              B\, {\rm ch\,}\tilde\eta_n \Bigr)
            {{{\partial}^2}\over {\partial}y^2}{\eta}_t(\tilde{\bf r}_{n})
\nonumber \\
    &+& \Bigl( A_n\, {\rm ch\,}\tilde\eta_n -
              B\, {\rm sh\,}\tilde\eta_n \Bigr)
        {\left({
         {{\partial}\over {\partial}y}{\eta}_t(\tilde{\bf r}_{n})
           }\right)}^2
      + B\, {{\rm sh\,}\tilde\eta_n \over \tilde x_n^2}\, ,
 \label{A6}\\
   {1 \over \lambda_o^2(\tilde x_n)}
   &=& \left( A_n\, {\rm sh\,}\tilde\eta_n -
              B\, {\rm ch\,}\tilde\eta_n \right) \tilde\eta_n''
     + \left( A_n\, {\rm ch\,}\tilde\eta_n -
              B\, {\rm sh\,}\tilde\eta_n \right) \tilde\eta_n'^2 \, .
 \label{A6a}
 \end{eqnarray}
Inserting for $\tilde x_n$ the saddle point $x^d_n$ of the
dynamical factor as determined by (\ref{A4}), the linear term in
(\ref{A5}) as well as the first terms on the right hand sides of
Eqs.~(\ref{A6},\ref{A6a}) vanish as they should:
 \begin{eqnarray}
  {1 \over \lambda_s^2(x^d_n)}
   &=& {B^2 \over \sqrt{A^2_n - B^2}} \,
       {1 \over (x^d_n)^2}\, ,
 \label{A6b}\\
   {1 \over \lambda_o^2(x^d_n)}
   &=& \sqrt{A_n^2 - B^2}\, (\eta^d_n)'^2 \, .
 \label{A6c}
 \end{eqnarray}
At any other point $\tilde x_n$ this no longer true. Inserting
the expansion (\ref{A5}) into (\ref{A2}) and completing the squares we
end up with a single gaussian
 \begin{equation}
   \ln G_n({\bf r}) \approx
    \ln C_n - {\delta  y^2 \over 2\, R_s^2(n)}
              - {\left(x - \bar x_n \right)^2
                 \over 2\ R_o^2(n) }\, ,
 \label{A8}
 \end{equation}
whose saddle point $\bar x_n$ is related to the point $\tilde
x_n$ around which the dynamical factor was expanded by
 \begin{equation}
   \bar x_n = \tilde x_n - \varepsilon_n
 \label{A7}
 \end{equation}
with
 \begin{equation}   \varepsilon_n = {R_o^2(n) \over R^2} \tilde x_n
   + R_o^2(n) \Bigl( A_n {\rm sh\,} \tilde \eta_n
                         - B {\rm ch\,} \tilde \eta_n \Bigr)
     \tilde \eta'_n \, .
 \label{A9}
 \end{equation}
The radius parameters in (\ref{A8}) are given in terms of the
geometric radius $R$ and the homogeneity lengths (\ref{A6},\ref{A6a})
by
 \begin{equation}
  {1\over R_s^2(n)}
   = {1\over R^2} + {1\over {\lambda}_s^2(\tilde x_n)}\, ,
 \qquad
   {1\over R_o^2(n)}
   = {1\over R^2} + {1\over {\lambda}_o^2(\tilde x_n)}\, ,
 \label{A10}
 \end{equation}
while the constant term is given by
 \begin{equation}
   \ln C_n = n + {1\over 2} - \Bigl( A_n \, {\rm ch\,}\tilde\eta_n
                            - B\, {\rm sh\,}\tilde\eta_n \Bigr)
                     - {\tilde x_n^2 \over 2 R^2}
                     + {\varepsilon_n^2 \over 2 R_o^2(n)} \, .
 \label{A11}
 \end{equation}
Identifying the expansion point $\tilde x_n$ with the true saddle
point $\bar x_n$ requires setting $\varepsilon_n{=}0$ and
amounts to solving the condition (\ref{18a}). The constant $J_n$
then simplifies accordingly. Instead of solving (\ref{18a})
numerically we can, however, solve (\ref{A4}) analytically for
$x^d_n$ and then set (see Eq.~(\ref{A6c}))
 \begin{equation}    \tilde x_n = { R^2 \over R^2 + \lambda_o^2(x^d_n) }
    \, x^d_n \, .
 \label{A12}
 \end{equation}
This is the saddle point of $G_n({\bf r})$ which would be obtained by
expanding $d_n({\bf r})$ around ${\bf r}^d_n$. We want to stress that
this procedure largely corrects for the fact that the radius
parameters $R^2_{s,o}(n)$ should be calculated from the curvature
at the full saddle point $\bar x_n$ instead of the saddle point
$x^d_n$ of the modified dynamical factor alone. As shown in
Section~\ref{sec5}, it is found to yield a very good approximation for
the HBT radii and their $m_\perp$-dependence.

\section{Calculation of HBT-radii}\label{appb}

In this appendix, we give expressions for the HBT-radii (\ref{12})
with coefficients calculated from the series expansions
(\ref{19}/\ref{20}) explicitly up to order $p=3$. Using the notational
shorthand
 \begin{equation}
    F_n = \left({T\over m_\perp}\right)^{n+{1\over 2}}
    R_o(n)\, R_s(n)\, C_n \, ,
 \label{B1}
 \end{equation}
and the results derived in Section~\ref{sec3B}, we obtain
 \begin{equation}
   R_s^2 = {\sum_{n=0}^p c_n\, F_n\, R_s^2(n)   \over
            \sum_{n=0}^p c_n\, F_n }\, ,
   \qquad
   c_n = \left( 1, \textstyle{3\over 8}, \textstyle{-{15\over 128}},
   \textstyle {105\over 1024}, \dots \right)\, ,
 \label{B2}
 \end{equation}
and
 \begin{equation}
  R_l^2 = \tau_0^2 {\sum_{n=0}^p \tilde{c}_n\, F_n  \over
            \sum_{n=0}^p c_n\, F_n }\, ,
   \qquad
   \tilde{c}_n = \left(0, 1, \textstyle{15\over 8}, \textstyle{105\over 128},
   \dots \right) \, .
 \label{B3}
 \end{equation}
The expansion of the ``out"-radius $R_o^2$ involves
three contributions:
 \begin{eqnarray}   \langle x^2 \rangle - \langle x \rangle^2
   &=& { \sum_{n=0}^p c_n\, F_n\,
         \left( R_o^2(n) + \bar x_n^2 \right) \over
         \sum_{n=0}^p c_n\, F_n }
     - { \left( \sum_{n=0}^p c_n\, F_n \, {\bar x}_n \right)^2
         \over
         \left( \sum_{n=0}^p c_n\, F_n \right)^2 }\,
 \nonumber \\
   &=& { \sum_{i,j=0}^p d^{xx}_{ij}\, F_i\, F_j\,
         \Bigl( R_o^2(i) + R_o^2(j) + (\bar x_i - \bar x_j)^2 \Bigr)
         \over
  \left( \sum_{n=0}^p c_n\, F_n \right)^2 }\, ,
 \label{B4}\\
   -2 {\beta}_{\perp}
   \Bigl( \langle xt \rangle - \langle x \rangle \langle t \rangle
   \Bigr)
   &=& \beta_\perp \tau_0
       { \sum_{i,j=0}^p d^{xt}_{ij}\, F_i\, F_j \, (\bar x_i - \bar x_j)
         \over
  \left( \sum_{n=0}^p c_n\, F_n \right)^2 }\, ,
 \label{B5}\\
   \beta_\perp^2 \left( \langle t^2 \rangle - \langle t \rangle^2 \right)
   &=& {\beta_\perp^2 \tau_0^2 \over 4}\,
        { \sum_{i,j=0}^p d^{tt}_{ij}\, F_i\, F_j \over
   \left( \sum_{n=0}^p c_n\, F_n \right)^2 } \, .
 \label{B6}
 \end{eqnarray}
Here, up to order $p = 3$ the coefficients are given by
 \begin{eqnarray}
   d_{ij}^{xx} &=& \left(
     \begin{array}{ccccc}
     \textstyle{1\over 2} & \textstyle{3\over 16} &
     \textstyle{-{15\over 256}} & \textstyle{105\over 2048} & ...  \\
     \textstyle{3\over 16} & \textstyle{9\over 128} &
     \textstyle{-{45\over 2048}} & ... & ...  \\
     \textstyle{-{15\over 256}} & \textstyle{-{45\over 2048}} & ... & ... & ...
\\
     \textstyle{105\over 2048} & ... & ... & ... & ... \\
     ... & ... & ... & ... & ...
     \end{array} \right) \, ,
 \label{B7}\\
   d^{xt}_{ij} &=& \left(
     \begin{array}{ccccc}
     0 & \textstyle{1\over 2} & \textstyle{9\over 16} &
     \textstyle{-{75\over 256}} & ... \\
     \textstyle{1\over 2} & 0 & \textstyle{69\over 256} & ... & ... \\
     \textstyle{9\over 16} & \textstyle{69\over 256} & ... & ... & ... \\
     \textstyle{-{75\over 256}} & ... & ... & ... & ... \\
     ... & ... & ... & ... & ...
     \end{array} \right) \, ,
 \label{B8}\\
   d^{tt}_{ij} &=& \left(
     \begin{array}{ccccc}
     0 & 0 & \textstyle{3\over 2} & \textstyle{45\over 16} & ...  \\
     0 & -1 & \textstyle{-{9\over 16}} & ... & ... \\
     \textstyle{3\over 2} & \textstyle{-{9\over 16}} & ... & ... & ... \\
     \textstyle{45\over 16} & ... & ... & ... & ... \\
     ... & ... & ... & ... & ...
     \end{array} \right)\, .
 \label{B9}
 \end{eqnarray}
It is interesting to note that only the expressions (\ref{B3}) and
(\ref{B6}) contain coefficients larger than $1$ (cf. our discussion in
Section~\ref{sec3C}).


\begin{figure}
\caption{
 $K_\perp$-dependence of the HBT radii $R_o$ (a), $R_s$ (b), and $R_l$
 (c), for the emission function (\protect\ref{8}) with parameters
 $T{=}150$ MeV, $\tau_0{=}3$ fm/$c$, $R$=3 fm. A linear transverse
 flow rapidity profile $\eta_t(r){=}\eta_{\rm f} (r/R)$ was assumed.
 Curves for different values of $\eta_{\rm f}$ are shown. The solid
 lines are calculated numerically from Eq.~(\protect\ref{4}); the
 long-dashed lines parametrize the widths $R_i^c$ of the numerically
 computed correlator $C({\bf K_{\perp}},q)$ according to
 Eq.~(\protect\ref{27}); the short-dashed lines represent our
 analytical results to order $p = 3$, while the dash-dotted lines
 denote the corresponding lowest order results.
}
\label{F1}
\end{figure}

\begin{figure}
\caption{
 Same as Fig.~\protect\ref{F1}, but for a quadratic transverse flow
 profile $\eta_t(r){=}\eta_{\rm f} (r^2/R^2)$. Only the exact HBT
 radii from a numerical integration of the model-independent
 expressions (\protect\ref{4}) (solid lines) and the width parameters
 $R_i^c$ of the numerically computed correlator $C({\bf K_{\perp}},q)$
 according to (\protect\ref{27}) (dashed lines) are shown.
}
\label{F2}
\end{figure}

\begin{figure}
\caption{
 Contour plots for the emission function (\protect\ref{10a}) in the
 transverse $x{-}y$ plane, with parameters $T{=}150$ MeV, $\tau_0{=}3$
 fm/$c$, $R$=3 fm, for a quadratic transverse flow profile
 $\eta_t(r){=}\eta_{\rm f} (r^2/R^2)$. From center to edge the lines
 correspond to 90\% \dots 10\% of the peak value (in steps of 10\%).
 The two left diagrams show that for a weak quadratic transverse flow
 with $\eta_{\rm f} = 0.1$) the emission region actually increases in
 the $x$-direction with increasing transverse momentum. The two right
 diagrams (for $\eta_{\rm f}=0.3$) show the generic decrease of the
 source in both $x$ and $y$ directions with increasing $K_\perp$.
}
\label{F3}
\end{figure}

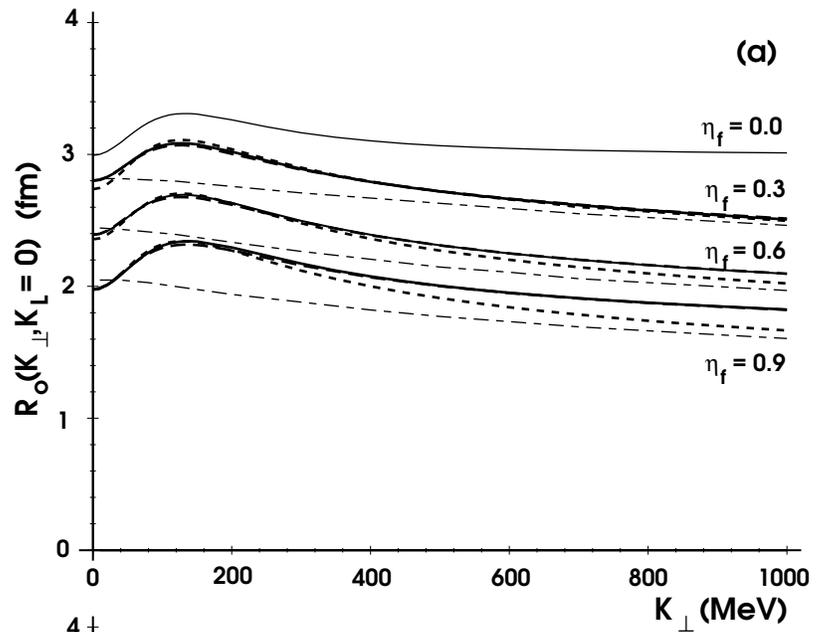
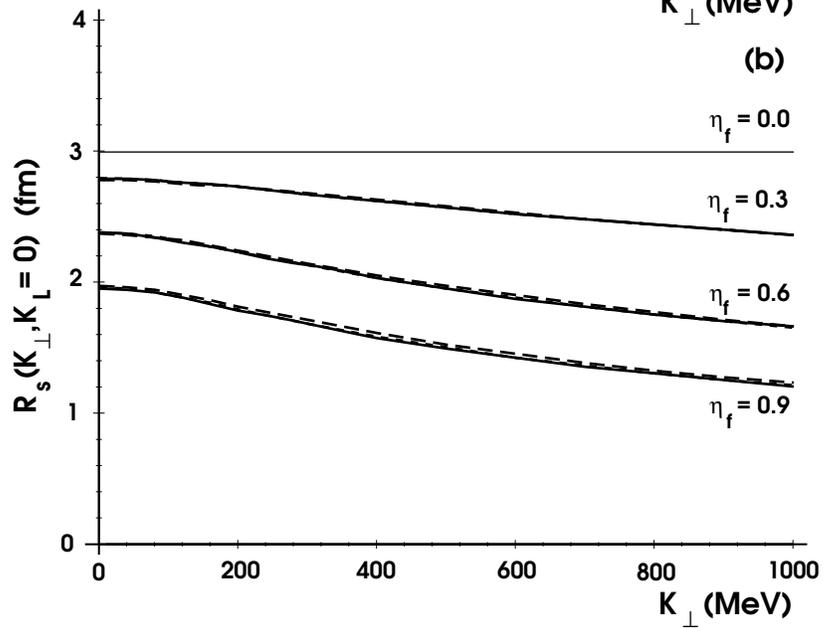
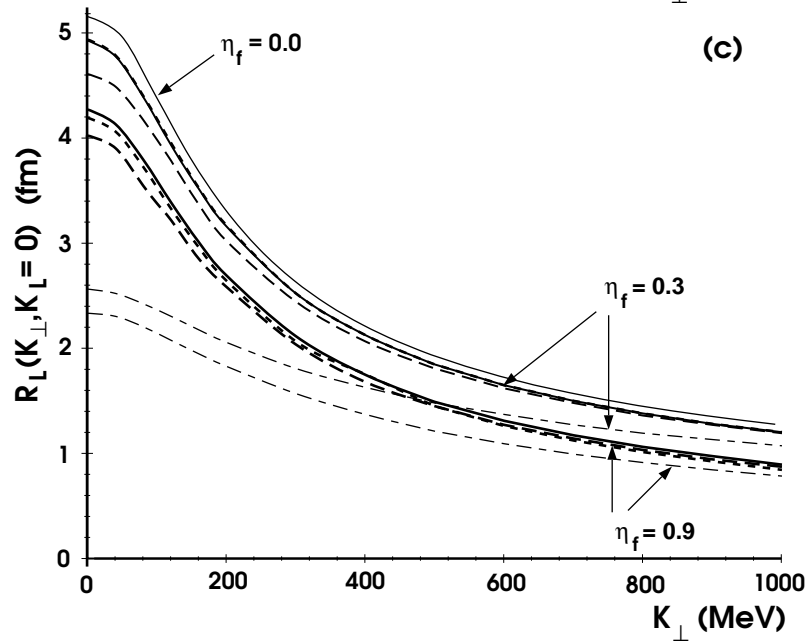
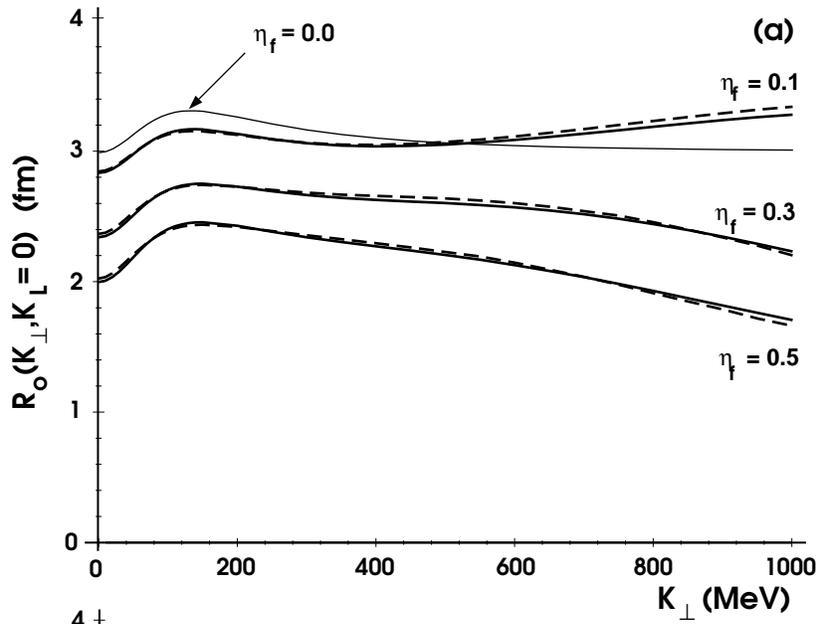
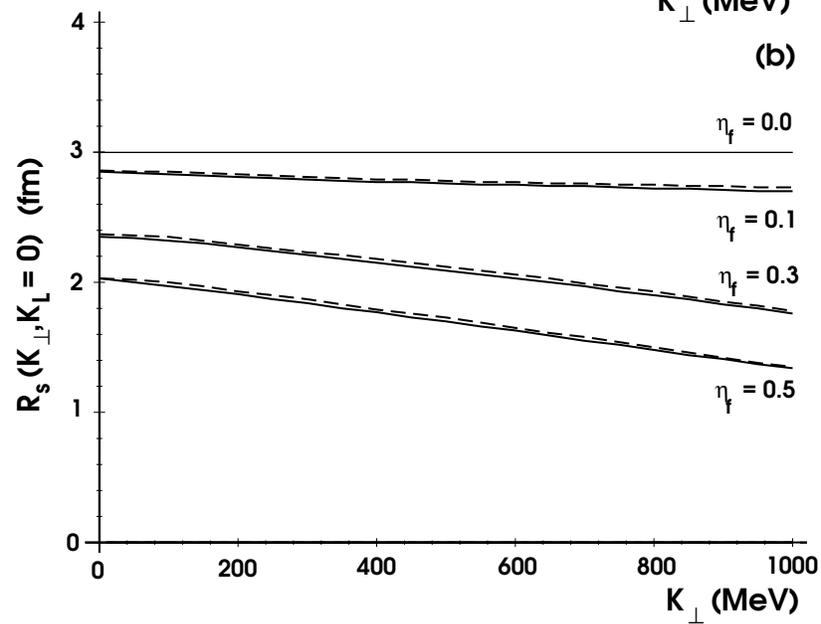
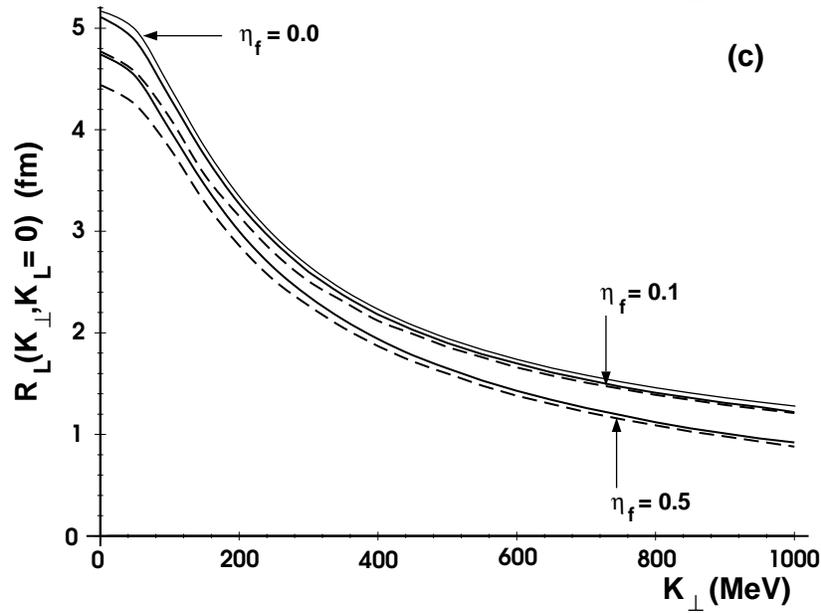
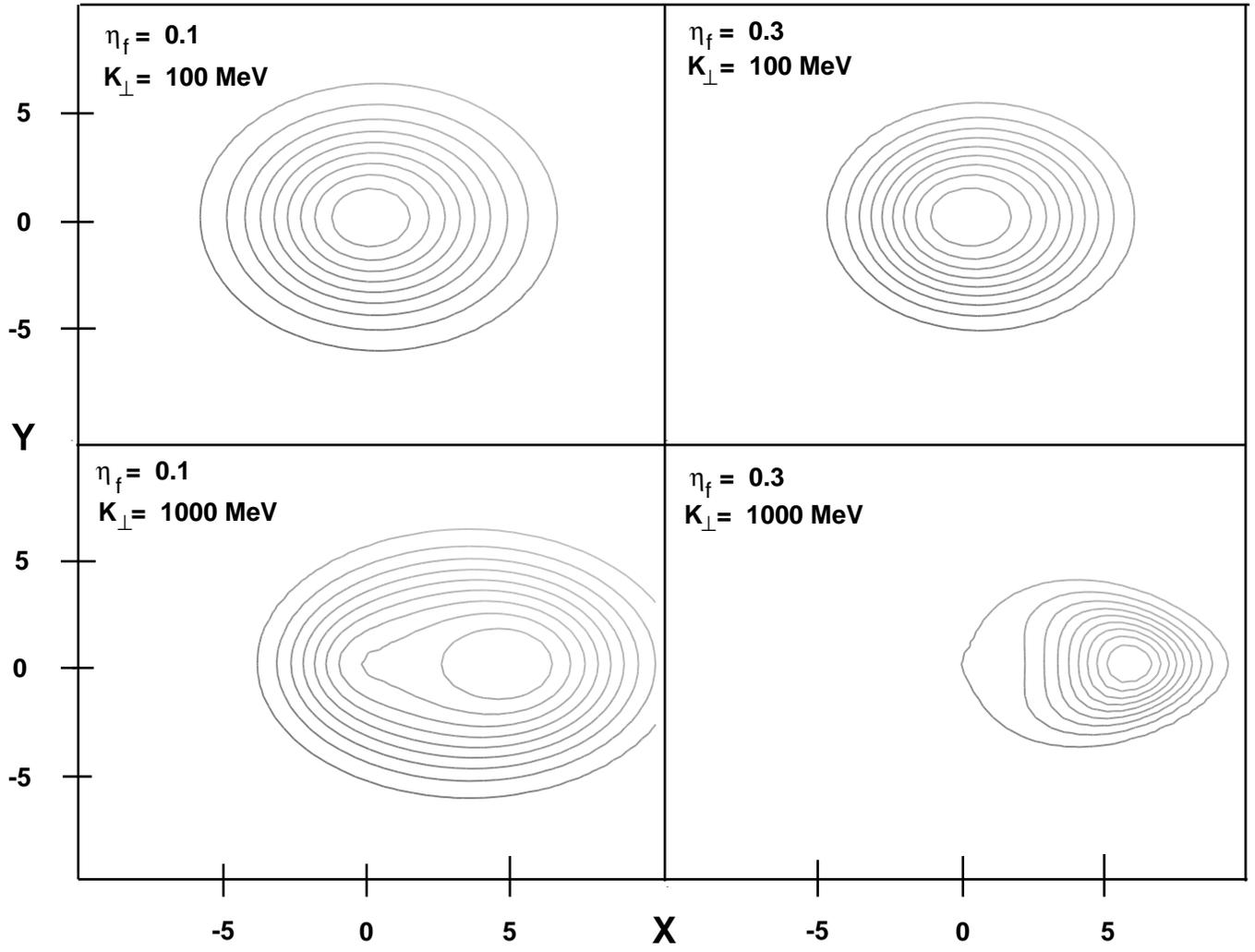
\end{document}